\documentclass{edm_article}
\usepackage{url}
\usepackage{svg}
\usepackage{graphicx}
\usepackage{booktabs} 
\usepackage{multirow} 

\usepackage{hyperref}

\begin{document}

\title{Towards Generalizable Representations of Mathematical Solution Strategies}

\numberofauthors{3}
\author{
Siddhartha Pradhan\\
\affaddr{Worcester Polytechnic Institute}\\
\email{sppradhan@wpi.edu}
\and
Ethan Prihar\\
\affaddr{Worcester Polytechnic Institute}\\
\email{ebprihar@wpi.edu}
\and
Erin Ottmar\\
\affaddr{Worcester Polytechnic Institute}\\
\email{erottmar@wpi.edu}
}

\maketitle

\begin{abstract}
    Pretrained encoders for mathematical texts have achieved significant improvements on various tasks such as formula classification and information retrieval. Yet they remain limited in representing and capturing student strategies for entire solution pathways. Previously, this has been accomplished either through labor-intensive manual labeling, which does not scale, or by learning representations tied to platform-specific actions, which limits generalizability. In this work, we present a novel approach for learning problem-invariant representations of entire algebraic solution pathways. We first construct transition embeddings by computing vector differences between consecutive algebraic states encoded by high-capacity pretrained models, emphasizing transformations rather than problem-specific features. Sequence-level embeddings are then learned via SimCSE, using contrastive objectives to position semantically similar solution pathways close in embedding space while separating dissimilar strategies. We evaluate these embeddings through multiple tasks, including multi-label action classification, solution efficiency prediction, and sequence reconstruction, and demonstrate their capacity to encode meaningful strategy information. Furthermore, we derive embedding-based measures of strategy uniqueness, diversity, and conformity that correlate with both short-term and distal learning outcomes, providing scalable proxies for mathematical creativity and divergent thinking. This approach facilitates platform-agnostic and cross-problem analyses of student problem-solving behaviors, demonstrating the effectiveness of transition-based sequence embeddings for educational data mining and automated assessment.
\end{abstract}


\section{Introduction}
Mathematical proficiency is an essential requirement for STEM disciplines and subsequent success in the STEM workforce\cite{grover_computational_2013}. Algebra, in particular, is considered foundational for learning more advanced topics in mathematics \cite{selden2009reflections}. However, many middle school and high school students struggle with basic algebraic concepts, including determining the validity of transitions, decomposing numbers, and converting simple algebra word problems to mathematically equivalent equations \cite{selden2009reflections, koedinger_real_2004}. 
Furthermore, students who struggle with these concepts may have difficulty learning more advanced STEM topics, most of which are usually represented or utilize symbolic and algebraic forms.

Across the globe, educational ministries and mathematics education researchers have highlighted strategy flexibility as an important facet in mathematics learning \cite{hong_systematic_2023}. To better understand more advanced mathematical concepts, students need to adaptively use multiple strategies to find appropriate and optimal solutions to math problems. Students' solutions to algebraic problems often unfold as a sequence of transitions, where the structure and ordering of intermediate procedural steps reflect students’ strategic choices.  

Although practically the Educational Data Mining community recognizes the student problem-solving process as a critical part of mathematical understanding, measuring these pathway-level characteristics in a problem- and platform-agnostic manner at scale remains an open challenge. While typical measures of problem performance, i.e., correctness, accuracy, or response time, capture whether and how quickly a student reaches a valid solution, process-based data on students' complete solution pathways may provide richer information about strategy efficiency, flexibility, and convergence in mathematical problem solving. Though many educational technologies can provide action-level data about students' behaviors during problem solving, prior works mostly focus on modeling platform-generated action sequences~\cite {thapa_magar_learning_2024, thapa_magar_can_2025} and learning representations of individual algebraic steps in a solution sequence~\cite{zhang_math_2021}. This limits the ability to compare strategies across learning environments, and learning representations of individual steps do not encapsulate the strategy of the entire solution. As a result, it is difficult to study the employed strategies in solution pathways without relying on expert-defined labels or manual coding.

To bridge this gap in research, we explore representation learning for \textit{entire} solution pathways that model algebraic problem-solving strategies. Specifically, we utilize high-capacity pretrained encoder models to embed raw states from a solution pathway, and aggregate the sequences of raw state embeddings using a SimCSE~\cite{gao_simcse_2021} based encoder to generate a fixed-vector representation of the entire solution pathway. To investigate the representational capacity of the resulting sequence-level embeddings, we evaluate them using a series of tasks that assess their ability to capture relevant sequence information.

Additionally, similar to prior work in linguistic creativity~\cite{beaty_automating_2021}, we use the resulting sequence-level embeddings to derive proxies for measuring creativity and divergent thinking based on semantic distance. Using prior data from a large Randomized Control Trial (RCT), these measures are then used to predict short-term and distal learning outcomes. Our results show that similarity measures of pathway-level representations capture meaningful variation in student strategies and are systematically associated with both short-term learning gains and longer-term achievement outcomes. 

Through our method in generating sequence-level embeddings and subsequent analyses, we contribute to the growing body of EDM research on analyzing student problem-solving behavior, offering a novel perspective on how the entire student solution sequences can reveal patterns of creativity and divergent problem-solving strategies.



\section{Related Work}
Learning representations for the entire solution pathways implies summarizing the employed strategies and all mathematical transitions. In prior work, Ritter et al. \cite{ritter_identifying_2019} discuss different approaches used to identify student strategies when solving math problems. Specifically, they define strategy as a sequence of steps or transitions taken to accomplish a given task or solve a problem. Additionally, they acknowledge that, while in theory, even slight variations in problem-solving represent different strategies, in practice, insignificant variations in problem-solving steps are grouped into a single strategy. Such a distinction requires a nuanced understanding of the intent of employed transitions. 

Studies, such as \cite{lee_perceptual_2022, baker2006human} have relied on the use of expert-labeled datasets to categorize student strategies and actions, which were created based on students' interactions in Intelligent Tutoring Systems or Digital Learning Platforms. While such datasets can be used to train machine learning models, i.e., detectors, it requires labor intensive human coding, which does not scale well to larger datasets. Additionally, strategy identification using human coding is generally subjective, depending on both the labeler and the context in which features are specified for the study \cite{paquette2014reengineering}. Thus, a more compact and generalizable representation of mathematical solution pathways is needed to support scalable and transferable analyses of student math problem-solving behaviors.

Alternatively, other studies have learned representations for mathematical sequences directly from log and interaction data. For example, \cite{shakya_student_2021} employed a Neuro-symbolic approach to encode student problem-solving strategies, which were defined as a sequence of knowledge components (i.e., KCs). Similarly, \cite{thapa_magar_can_2025, thapa_magar_learning_2024} trained a small BERT variant to encode sequences of actions specific to the MATHia ITS (e.g. ``DenominatorFactor", ``FinalAnswer-1"). Other studies focused on a subset of strategies available in the log data and developed methods to identify and differentiate strategies that led to mathematical errors\cite{feldman_automatic_2018, kolb_generalizing_2019}.


Overall, these approaches either focused on identifying specific error-related strategies or relied on predefined sequence labels such as KCs or system-specific actions. This limits their generality and transferability across problems and systems.

\subsection{Learning Representations From Math Formulas} \label{section-formula-emb}
Another related body of work have learned representations directly from mathematical formulas. For example, MathBERT~\cite{shen_mathbert_2023} further pre-trained the BASE BERT model on mathematical domain texts, ranging from pre-K to graduate-level mathematical curricula, including formulas and their contexts. The model demonstrated the ability to perform ``general" mathematics-related tasks such as KC prediction or open-ended question answer scoring. Zhang et al.~\cite{wang_scientific_2021} took a different approach, where they focused solely on the formula and disregarded additional natural language context. Specifically, they represented a formula as a tree structure and used an autoencoding model design to learn its representation. This method was later used in~\cite{zhang_math_2021} to predict the transitions (i.e., add, multiply, distribute) between math expressions in consecutive steps. While these various approaches focused on embeddings of contextualized formulas or individual equation states, they did not directly address representation learning for the strategies and transitions for the \textit{entire} solution pathway. Our work complements and builds on this line of research by explicitly modeling transitions and composing them into pathway-level representations. 

\subsection{Automated Measurement of Creativity} \label{section-auto-creativity}
While prior work on mathematical embeddings has primarily focused on downstream tasks such as formula headline generation and KC prediction, an open question is whether distances in learned embedding spaces can be used to characterize latent constructs in mathematical problem solving such as strategy diversity or strategic flexibility. Related work in the creativity literature offers a complementary perspective, demonstrating that semantic distance in embedding spaces can serve as a scalable proxy for complex cognitive constructs like strategy flexibility. 

Early measures of creativity and divergent thinking often relied on human labeling to code strategies~\cite{kwon_cultivating_2006, leikin_exploring_2009}, which is labor-intensive, subjective, and difficult to scale. However, to address these limitations, more recent approaches have explored the use of automated methods based on semantic similarity. For example, in \cite{beaty_automating_2021}, the authors present the SemDis tool, which scores responses to Unusual Uses Tasks (UUTs) that measure divergent thinking. To assess variations in creativity, the tool computes the semantic distance between a target object (e.g., brick) and a participant’s responses (e.g., weapon, wall) using cosine similarity. Greater semantic distance reflects more remote associations and higher creative potential. These results demonstrate that embedding-based similarity measures can function as scalable proxies for cognitive constructs traditionally assessed through human coding. 

\section{Current Work}
Taken together, while numerous advancements have been made in prior work, there is a need to identify new methods for characterizing and analyzing students' complete solution pathways when solving mathematics problems. To address these gaps, we explore the following three research questions:
\begin{itemize}
    \item[\textbf{RQ 1}] How do transition-based embeddings compare to state-based embeddings in capturing the general structure and semantics of mathematical transitions?
    \item[\textbf{RQ 2}] Do contrastively learned sequence embeddings of full solution pathways capture meaningful representations of employed problem-solving strategies?
    \item[\textbf{RQ 3}] How are embedding-derived measures of pathway uniqueness, diversity, and conformity associated with short-term and long-term student learning outcomes?
\end{itemize}

\section{Background}
\subsection{Facets of Algebraic Learning}\label{sec:algebraic-learning}
Algebraic proficiency requires not only procedural and conceptual knowledge, but also the ability to apply this knowledge flexibly \cite{schneider_relations_2011}. Procedural knowledge is defined as the ordering of algebraic steps and transitions required to solve a problem, whereas conceptual understanding relates to the knowledge of core algebraic concepts such as symbols and syntactic conventions \cite{rittle-johnson_developing_2001}. Both types of knowledge complement the other: conceptual knowledge enables developing procedural knowledge \cite{rittle-johnson_not_2015}, and consistent procedural practice reinforces conceptual knowledge \cite{rittle-johnson_developing_2001, schoenfeld_problem_2007}.

Both conceptual and procedural knowledge facilitate math flexibility, defined as the capacity to use multiple solution strategies and subsequently select the most efficient or appropriate for a given problem \cite{rittlejohnson_developing_2012}. While flexibility has been internationally recognized as a core goal for math instruction, encouraging it in practice and measuring it in research has been challenging \cite{hong_systematic_2023}. Although several studies suggest the benefit of encouraging students to compare and contrast different strategies, instruction often focuses on a single preferred method \cite{verschaffel_conceptualizing_2009, alfieri2013learning, stanton2021fostering}. Additionally, exposure to multiple strategies allows students to recognize and select when particular strategies are valid and applicable \cite{knuth_does_2006, ottmar_teaching_2012, welder_improving_2012}.  Students' conceptual knowledge, procedural understanding, and mathematical flexibility are all critical aspects of mathematical understanding. Yet, few studies have examined how the sequences of students' problem-solving strategies relate to these three facets of mathematical knowledge.

\subsubsection{Strategy Flexibility and Math Creativity}
These facets of algebraic learning closely align with core components of creative problem-solving. In particular, flexibility in generating among multiple solution pathways or strategies reflects the cognitive processes underlying more divergent thinking, which emphasizes generating varied and more creative original solutions. On the other hand, convergent thinking pushes students towards a single solution or reflects more common or efficient pathways. An important question in mathematics education is whether divergent or more convergent solution pathways (or both) support students math learning. 

Mathematical creativity, in the context of problem-solving, can be considered as a combination of originality, fluency, and flexibility expressed in solving a math problem \cite{leikin2007multiple}. Originality is assessed based on how unconventional and insightful a problem’s solution method is, while fluency is measured by the number of distinct, non-repeating solution methods a student produces for a given problem \cite{leikin2007multiple, tabach_algebraic_2017}. 

Similar to the use of the SemDis tool to measure linguistic creativity (see Section~\ref{section-auto-creativity}), learning representations of student math solution pathways may allow us to quantify originality, fluency, and flexibility in algebraic problem solving. Distances within this learned space can capture the diversity of a student's strategies and the novelty of their solution relative to peers, providing a scalable measure of creative problem-solving behavior.

\subsection{Vector Representation for Entire Sequences} \label{section-sentence-emb}
Section~\ref{section-formula-emb} discussed recent advancements in learning vector representations for mathematical contexts and individual formulae, typically producing embeddings $\mathbf{h}_t \in \mathbb{R}^d$ for each state or formula $s_t$ within a sequence. While these representations capture local structure and semantic information for a single state, they do not provide a representation of the entire solution sequence $\mathbf{S} = (s_1, s_2, \dots, s_T)$, which is essential for characterizing global solution strategies and overall patterns across multiple steps. Formally, the challenge is to define a mapping $f: \mathbf{S} \mapsto \mathbf{z}_S \in \mathbb{R}^k$ that embeds the full sequence $\mathbf{S}$ into a latent space $\mathbb{R}^k$ in a manner that preserves important transitional and structural information.

The task of embedding entire sequences is commonly referred to as sentence embedding in the natural language processing literature. A straightforward approach is to aggregate step-level embeddings across the sequence dimension using operations such as averaging or max pooling. However, these aggregation-based methods often fail to capture higher-order sequential structure and semantic dependencies. Consequently, more advanced approaches have been proposed to learn sequence-level representations directly \cite{hill_learning_2016, logeswaran_efficient_2018, reimers_sentence-bert_2019, gao_simcse_2021}. Motivated by these NLP approaches, we explore sequence-level embeddings for mathematical problem-solving, seeking representations that encode the structure and strategy in students’ solution pathways.

\section{Methods}
\subsection{Data and Context}
In this section, we describe the overall context of the study, including the learning platform from which the data was obtained and its source.

\subsubsection{From Here To There! (FH2T)}
FH2T is a gamified learning platform grounded in principles of perceptual learning, embodied cognition, and gamification \cite{ottmar_getting_2015}. The design of FH2T emphasizes the structural properties of algebraic expressions and encourages students to reason flexibly about mathematical operations rather than relying on fixed procedural steps \cite{chan_slow_2022}. Prior research has shown that FH2T leads to significant improvements in students’ understanding of equivalence, mathematical flexibility, and both conceptual and procedural knowledge \cite{chan_slow_2022, decker-woodrow_impacts_2023, ottmar_getting_2015, pradhan2025mathflowlens}.  

The FH2T game has 252 total problems and is organized into 14 ``worlds" corresponding to mathematical concepts such as addition and multiplication. Within each world, problems are ordered by increasing difficulty, and students need to complete all preceding problems before advancing to the next world \cite{lee_does_2022}. Solving a problem rewards students based on the efficiency of their solution strategy.

\begin{figure}[ht]
    \Description{Example Problem 7 from FH2T with sample solution steps (Start State: 11+55+y+89+45, Goal State: 100+y+100}
    \centering
    \includegraphics[width=\linewidth]{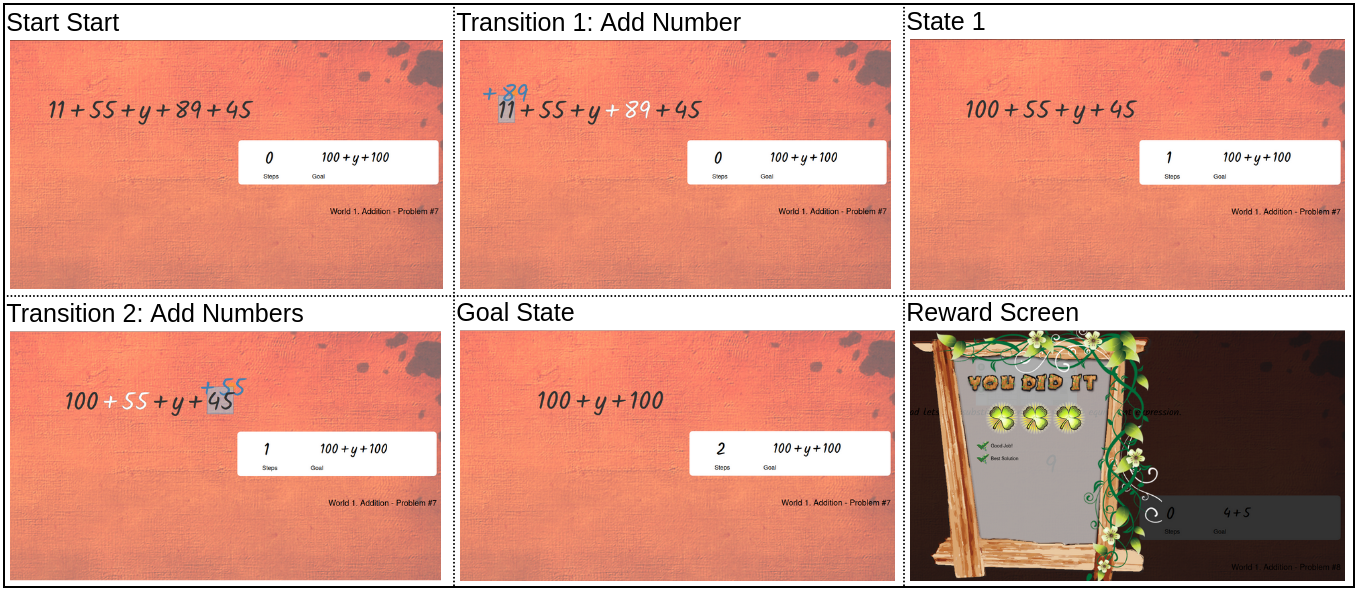}
    \caption{Example Problem 7 from FH2T with sample solution steps (Start State: $11+55+y+89+45$, Goal State: $100+y+100$}
    \label{fig:fh2t_example}
\end{figure}

In FH2T, algebraic equations are represented as interactive virtual objects that students manipulate through dynamic actions such as tapping and dragging. Each problem begins with an initial algebraic expression, referred to as the start state, which must be transformed into a mathematically equivalent but perceptually different target expression, referred to as the goal state (see Fig.~\ref{fig:fh2t_example}). For example, the expression ``$11+55+y+89+45$'' may be transformed into ``$100+y+100$'' through a sequence of valid algebraic transitions. One possible solution path involves first adding ``$11$'' and ``$89$'' to obtain the intermediate state ``$100+55+y+45$'', followed by adding ``$55$'' and ``$45$'' to reach the goal state ``$100+y+100$''.

\begin{figure}[h]
    \Description{Multiple solution strategies identified by a student for Problem 7. All three solutions use different strategies, resulting in different efficiencies, employed transitions, and explores math states.}
    \centering
    \includesvg[width=0.7\linewidth]{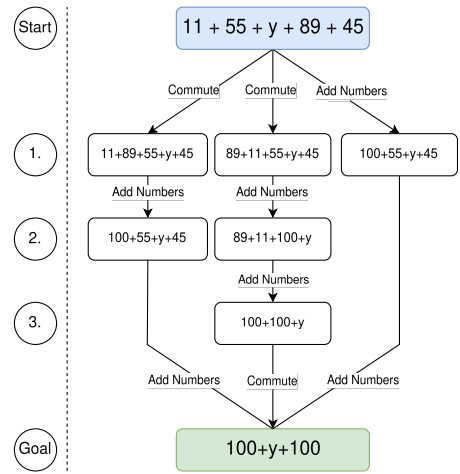}
    \caption{Multiple solution strategies identified by a student for Problem 7. All three solutions use different strategies, resulting in different efficiencies, employed transitions, and explores math states.}
    \label{fig:fh2t_strats}
\end{figure}

FH2T is an open-ended gesture-based platform where students are free to choose any procedural pathway they prefer, with all mathematically valid transitions being permissible. Additionally, students are allowed to reset a problem to its starting state at any point during problem solving and may also replay problems after completion. This freedom inherently grants students the autonomy to explore different solution strategies and various transitions required to complete a problem \cite{pradhan_gamification_2024}. Figure~\ref{fig:fh2t_strats} displays sample solutions strategies explored by a student in the FH2T logs. Each of their strategies varies in efficiency, as well as the employed math transitions and state sequences. Hence, the logged data collected from FH2T allows us to analyze diversity in strategy, flexibility, and pathway-level problem-solving behaviors.

\subsubsection{Data Source}
To learn representations of students’ problem-solving strategies and relate them to learning outcomes, we used data from a large-scale randomized controlled trial (RCT) of FH2T conducted during the COVID-19 pandemic between September 2020 and April 2021. A total of 4,092 seventh-grade students from 11 middle schools in a large suburban school district in the United States participated in the study \cite{ottmar_data_2023}. The primary outcomes of the RCT are reported in \cite{decker-woodrow_impacts_2023}. The dataset is publicly available through the Open Science Framework (OSF)\footnote{Data is available through \url{https://osf.io/r3nf2/}.} as described by the dataset paper \cite{ottmar_data_2023}.

The assessment data of the RCT includes pre-test and post-test measures administered before and after the intervention. Each assessment consists of 10 items, measuring conceptual knowledge (4 items), procedural knowledge (3 items), and mathematical flexibility (3 items). These measures were adapted from previously validated instruments \cite{ottmar_data_2023}. In addition, the dataset includes standardized state mathematics assessment scores for fifth and seventh grade that were provided by the district. Sixth-grade state assessment scores are unavailable due to pandemic-related disruptions.

Of the 4,092 students in the full RCT sample, 1,649 were assigned to the FH2T condition. Within this group, 52.6\% of students identified as male and 47.4\% as female. The racial and ethnic composition was 49.8\% White, 24.8\% Asian, and 16.4\% Hispanic or Latino, with the remaining 9.0\% identifying with multiple racial categories \cite{decker-woodrow_impacts_2023}. Among the FH2T cohort, 1,108 completed the pre-test, and 777 completed the post-test. These students completed an average of 111 problems (\textit{SD} = 55.05) out of a total 252 problems in the FH2T game, generating the algebraic problem-solving sequences used in the current study. This fine-grained log data from the FH2T algebra game enabled detailed analysis of students' solution sequences and strategies.


\subsubsection{Data Preprocessing}
The FH2T log data collected in the RCT contain a total of 70,257 unique algebraic states and 128,050 distinct state-to-state transitions across 252 problems. These states and transitions result in 193,782 problem-solving pathways, many of which are repeated by multiple students solving the same problem. To mitigate potential biases during model training, duplicate or extremely long sequences were removed. Specifically, sequences exceeding the 0.95 quantile length, corresponding to 15 steps, were discarded, resulting in a maximum sequence length of 15. After this filtering, a total of 81,306 sequences remained for analysis.

Each transition in the dataset is associated with one of 75 distinct ground truth action labels (e.g., AddSubNumbers, MultiplyNumbers, ExpressionRewrite, Commute, etc.) corresponding to FH2T's internal classification of algebraic transformations. Considering the sequences of these action labels, there are 28,573 unique label sequences. However, as our goal is to produce sequence embeddings that are agnostic to the platform, the action labels are unused during model training. 

\subsection{Learning  Representations Entire Solution Sequences}
The methods described in Section~\ref{section-formula-emb} produce embeddings for individual formulas or equation states rather than complete solution sequences. Nevertheless, these models were pretrained on large corpora of mathematical expressions and text, and have demonstrated strong performance across a range of math-related tasks, indicating substantial representational capacity. We therefore used these pretrained models as fixed encoders for individual states and used their outputs to obtain representations of entire solution sequences using sequence-level objectives. A summary of the pretrained encoders used in this study is provided in Table~\ref{tab:state-encoders}.


\begin{table}[t]
\centering
\caption{Pretrained encoders used as fixed state encoders.} 
\label{tab:state-encoders}
\begin{tabular}{lll}
\hline
HuggingFace ID & $d_{model}$ & Reference \\
\hline
\texttt{AnReu/math\_pretrained\_bert} & 768 & \cite{reusch2022transformer, reusch_investigating_2024} \\
\texttt{tbs17/MathBERT} & 768 & \cite{shen_mathbert_2023} \\
\texttt{aieng-lab/MathBERT-mamut} & 768 &  \cite{drechsel_mamut_2025} \\
\hline
\end{tabular}
\end{table}

\subsubsection{Comparing State vs.\ Transition Embeddings}
In most representation learning pipelines, embeddings for a sequence are directly aggregated from various layers of a model to form sentence- or sequence-level representations~\cite{ma_universal_2019}. This approach is effective when the objective is to preserve fine-grained similarity between highly similar sequences. However, in the context of identifying and comparing algebraic problem-solving strategies, such methods may encode detailed state-level structures that are specific to the problem formulation rather than the employed strategies. In other words, two solutions that use the same strategy (e.g., adding two numbers) but differ in mathematical content (e.g., $x+1+1 \rightarrow x+2$ and $y+5+5 \rightarrow y+10$) may be far apart in the embedding space. This may limit generalizability across problems and platforms. 

To address this concern, we investigated two ways to represent a student's solution pathway: \emph{state-based} embeddings and \emph{transition-based} embeddings. Let a student’s solution pathway be represented as an ordered sequence of algebraic states $\mathbf{S} = (s_1, s_2, \dots, s_T)$, and let $g_\theta$ denote one of the pretrained encoders listed in Table~\ref{tab:state-encoders}. State-based embeddings map each state $s_t$ to a vector representation: $\mathbf{H} = (h_1, h_2, \dots, h_T)$, where $\mathbf{h}_t = g_\theta(s_t)$. For transition-based embeddings, we draw inspiration from the translating embedding framework (TransE) \cite{bordes_translating_2013}, which has previously been shown to be effective for modeling mathematical transitions \cite{zhang_math_2021}. A core assumption in this framework is that math transitions are \textit{linear} and \textit{additive} in the embedding space, i.e., $\delta_t = h_t - h_{t-1}$. Hence, the transition-based embeddings is given by: $\mathbf{\Delta} = (\delta_2, \delta_3, \dots, \delta_{T})$.

To assess which method of representing solution sequences better captures solution strategies, we compared the geometric structure of state- and transition-based embeddings. Specifically, we applied t-SNE~\cite{maaten2008visualizing} to project the embeddings into a two-dimensional space for qualitative analysis of their clustering and separation patterns. We restricted this analysis to a subset of the data consisting of the ten problems with the longest solution sequences and retained only states and transitions corresponding to the ten most frequent action types in the dataset.

\subsubsection{Model Architecture}

The methods above produce $d$-dimensional embeddings for each state or transition at each time step within a solution sequence. For our chosen pretrained encoders, both state and transition embeddings have dimension $d=768$. These form sequences $\mathbf{H} \in \mathbb{R}^{T \times d}$ or $\boldsymbol{\Delta} \in \mathbb{R}^{(T-1) \times d}$ that is used as input for the model. 

We encoded the full solution sequence using a Transformer-based model \cite{vaswani2017attention} that contextualizes the input embeddings and produces a fixed-dimensional sequence representation $\mathbf{z} \in \mathbb{R}^k$, where $k$ is a hyperparameter denoting the sequence-embedding dimension. In the model, each state or transition embedding is first mapped into a higher-dimensional space using a dense layer with Gaussian Error Linear Units (GELU) activation \cite{hendrycks_gaussian_2023}, followed by dropout \cite{srivastava_dropout_2014} and a projection back to the embedding dimension $k$. Positional information was added via a learned position embedding \cite{vaswani2017attention}.

\begin{figure}[!h]
    \Description{Illustration of the SimCSE-based encoder model, which uses Transformer layers as the backbone. The model takes as input embeddings produced by a frozen encoder for individual states or transitions in a solution pathway and outputs sequence-level representations.}
    \centering
    \includesvg[width=0.5\linewidth]{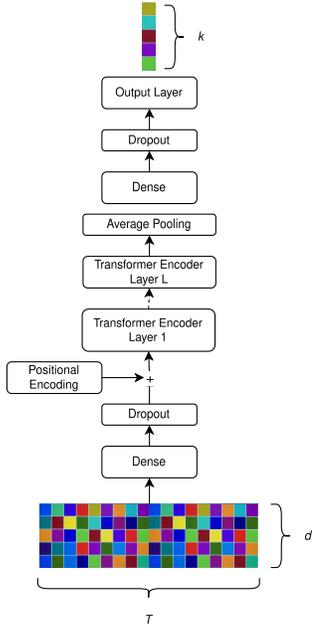}
    \caption{Illustration of the SimCSE-based encoder model, which uses Transformer layers as the backbone. The model takes as input embeddings produced by a frozen encoder for individual states or transitions in a solution pathway and outputs sequence-level representations.}

    \label{fig:transformer_model}
\end{figure}

The core of the model consists of $L$ stacked Transformer encoder layers \cite{vaswani2017attention}, each with multi-head self-attention and feed-forward sublayers, where the number of transformer layers $L$ is a hyperparameter. After the Transformer stack, the sequence of contextualized vectors is aggregated using global average pooling, followed by a feed-forward hidden layer with dropout and a final projection to the embedding dimension (see Figure~\ref{fig:transformer_model}). 

\subsubsection{Model Training}
 
While several approaches have been proposed for aggregating sequences into fixed-dimensional representations (see Section~\ref{section-sentence-emb}), we employed SimCSE \cite{gao_simcse_2021}, due to its efficient training procedure and state-of-the-art performance in producing high-quality sentence embeddings in an unsupervised setting. 

SimCSE is a contrastive learning framework that explicitly encourages semantically similar sequences (positives) to have closer embeddings while pushing apart dissimilar sequences (negatives). Let $\mathbf{z}_S = f_\theta(\mathbf{X}) \in \mathbb{R}^k$ denote the embedding of a full solution sequence, where $\mathbf{X}$ is either the state-based ($\mathbf{H}$) or the transition-based ($\boldsymbol{\Delta}$) embedding sequence, and $f_\theta$ is our proposed encoder model. Positive pairs are constructed by applying independent dropout masks to the same sequence, resulting in embeddings $(\mathbf{z}_S, \mathbf{z}_S')$ that are considered semantically equivalent. Negative pairs consist of embeddings of other sequences within the mini-batch, $(\mathbf{z}_{S_i}, \mathbf{z}_{S_j}')$ where $i \neq j$. The model is trained with the objective:

\[
\mathcal{L}_S = - \log \frac{\exp(\mathrm{sim}(\mathbf{z}_S, \mathbf{z}_S') / \tau)}{\sum_{S' \in \mathcal{B}} \exp(\mathrm{sim}(\mathbf{z}_S, \mathbf{z}_{S'}) / \tau)},
\]

where $\mathrm{sim}(\mathbf{{z}_1, {z}_2})$ denotes cosine similarity, $\tau$ is a temperature hyperparameter, and $\mathcal{B}$ is the set of sequences in the mini-batch. By optimizing this objective, semantically equivalent sequences occupy nearby regions while unrelated sequences are separated in the embedding space. 

For each frozen pretrained encoder $g_\theta$, we train our model over a grid of hyperparameters, varying the number of Transformer layers $L \in \{1,2,4\}$ and the sequence-embedding dimension $k \in \{128,256,512,768\}$. This results in a total of $3 \times 3 \times 4 = 36$ configurations, where each model is trained with early stopping based on validation loss.

\subsection{Evaluation of Sequence Level Embeddings}
In NLP literature, sentence embeddings are typically evaluated through benchmarks: human judgment of sentence similarity \cite{agirre2016semeval, conneau_senteval_2018} or performance in select downstream tasks \cite{kennard2016evaluating}. However, such human-labeled datasets or specific downstream tasks are not available for student problem-solving strategy representations. To evaluate the representational capacity of the sequence embeddings, we investigated the types of information captured by these representations. Specifically, similar to \cite{adi_fine-grained_2017} that used auxiliary tasks to analyze sentence embeddings, we trained small probes to evaluate performance on tasks specific to the sequence. Having good performance on these tasks may indicate that our embeddings captured the employed solution strategies. We propose the following three tasks based on the available data and prior literature:

\begin{enumerate}
    \item \textbf{Action Type content:} In FH2T, each transition in the solution pathway is classified as 1 of 75 system action types. This task measures the extent to which the sequence representation encodes the identities of action types within it. Since a solution contains multiple transformations and hence action types, this is a multi-label classification task. This is similar to the work in~\cite{zhang_math_2021}, where embeddings from two consecutive states were used to predict the actions. However, we predict the presence or absence of \textit{all} actions in a solution sequence.
    \item \textbf{Efficiency of the solution:} Solution pathways for a given problem have varying levels of efficiency. Some strategies are more effective and direct, which lead to shorter pathways. Each problem in FH2T has a minimum number of actions required to solve it. Based on this, we categorized each unique solution pathway as \emph{optimal}, \emph{suboptimal}, or \emph{incomplete} (i.e., the student reset the problem before completion). Prior work has shown that students with a higher frequency of efficient solution pathways are associated with improved learning outcomes \cite{pradhan_gamification_2024, pradhan2025mathflowlens}. Consequently, the ability to distinguish inefficient or incomplete strategies is a critical property of effective pathway representations. 
    \item \textbf{Action Sequence Reconstruction:} While the Action Type content tests whether the embeddings encode the identities of the actions in the sequence, the order of the actions plays a critical role in determining the solution strategy. Prior works have explored embedding inversion~\cite{li_sentence_2023, morris_text_2023}, where sentence embeddings are passed to a decoder to auto-regressively reconstruct the original sentence. We propose a similar task, where we used the sequence embeddings to reconstruct the original action sequence. This task serves to assess whether the order of actions is preserved.  
\end{enumerate}

We fitted an MLP with a single hidden layer and a GELU nonlinearity for both Action content and Efficiency classification tasks. However, these probes differ in output dimensions and training procedure: the action content probe has a 75-dimensional output layer with sigmoid activation and was trained using binary cross-entropy, whereas the efficiency prediction probe has three output units with softmax activation and was trained using categorical cross-entropy. Unlike the MLP probes, the Action Sequence reconstruction task requires a more complex probe due to its auto-regressive nature. We implemented the probe as a single-layer LSTM~\cite{hochreiter_long_1997}, and the input sequence embeddings are mapped to the initial hidden and cell states using an MLP with Tanh activation. This architecture enables step-by-step sequence generation while conditioning on the information encoded in the learned embeddings.

We trained each probe type for all model configurations (i.e., for each frozen encoder and corresponding set of hyperparameters), and used early stopping based on validation loss. Additionally, for a baseline comparison, we fit probes on the mean-pooled embeddings from the frozen encoders. We assessed the best-performing configuration and the representational capacity of the resulting embeddings using relevant performance metrics. For classification-based tasks, we used micro- and macro-averaged F1 scores, while for sequence reconstruction, we used perplexity and token-level accuracy.

\subsection{Similarity and Distance Based Strategy Measures}
In Sections~\ref{section-auto-creativity} and \ref{sec:algebraic-learning}, we reviewed prior work on automated and scalable assessment of creativity via semantic distance \cite{beaty_automating_2021}, as well as theoretical constructs of creativity and their relationship to different aspects of algebraic learning. Here, we define potential proxies for measuring creativity based on originality, fluency, and flexibility.

\begin{enumerate}
    \item \textbf{Strategy Uniqueness} reflects how unconventional and unique a student’s solution methods are relative to other students in the cohort. We operationalized this as the Mahalanobis distance between a student’s solution-sequence embedding and the distribution of the embeddings across all student attempts in FH2T. Distances were computed per attempt and averaged to obtain a student-level originality score, where larger values indicate that a student’s solution sequences deviate more from commonly used cohort-level strategies.

    \item \textbf{Strategy Diversity} captures the number of distinct solution methods a student produces for a given problem \cite{leikin2007multiple, tabach_algebraic_2017}. 
    For each student, we computed the pairwise cosine \textit{dissimilarity} (i.e., 1 - cosine-similarity) between all of their solution attempts and took the average of the lower-triangular entries of the resulting dissimilarity matrix. Higher strategy diversity corresponds to higher fluency and more exploration in the solution space.

    \item \textbf{Strategy Conformity} measures a student’s ability to identify and apply common and effective strategies. For each problem, we computed the cosine similarity between a student’s solution embedding and the embedding for the most frequently observed optimal solution for that problem. These similarities were first standardized within each problem and then averaged to obtain a student-level measure, where higher values indicate closer alignment with widely used optimal strategies.
\end{enumerate}

Using these measures, we fitted regression models to assess the relationship between the proposed creativity proxies and student learning outcomes. Specifically, we examined how strategy uniqueness, diversity, and conformity are associated with distal outcomes such as state test scores, as well as short-term learning metrics including conceptual understanding, procedural skills, and flexibility. This analysis enables an evaluation of how distinct dimensions of creative problem-solving are associated with observed learning outcomes.

\section{Results}

\subsection{RQ1: Generalizability of State vs. Transition Embeddings}

\begin{figure*}[!h]
    \Description{t-SNE projections of individual state- and transition-level embeddings. 
                (a) State-based embeddings colored by problem ID. 
                (b) Transition-based embeddings colored by problem ID. 
                (c) Transition-based embeddings colored by action type.}
    \centering
    \includesvg[width=0.7\linewidth]{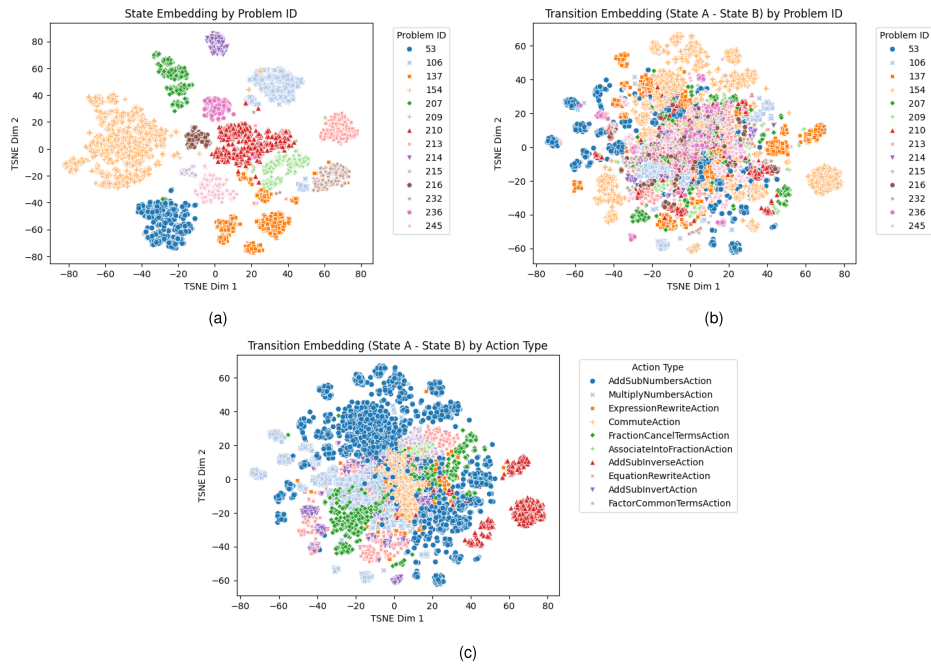}
    \caption{t-SNE projections of individual state- and transition-level embeddings. 
            (a) State-based embeddings colored by problem ID. 
            (b) Transition-based embeddings colored by problem ID. 
            (c) Transition-based embeddings colored by action type.}
    \label{fig:tsne_embeddings}
\end{figure*}

Figure~\ref{fig:tsne_embeddings} visualizes the geometric structure of \emph{individual} state- and transition-based embeddings using t-SNE projections. Each point corresponds to a single algebraic state or transition. For state-based embeddings (Figure~\ref{fig:tsne_embeddings}a), the embeddings form well-separated clusters that align closely with their corresponding problem ID. This indicates that state embeddings retain substantial problem-specific information, likely due to the surface form of algebraic expressions. Consequently, states from different problems are located far apart, even when they may correspond to similar algebraic operations.

In contrast, the transition-based embeddings in Figure~\ref{fig:tsne_embeddings}b showed no apparent clustering. Instead, these transition embeddings exhibited substantial overlap across problems, indicating invariance to problem-specific content. However, when transition embeddings were viewed by their corresponding action type (Figure~\ref{fig:tsne_embeddings}c), clearer clustering patterns emerged that corresponded to the mathematical transition (e.g., addition, multiplication, factorization). This organization indicated that transition embeddings are structured primarily around the type of algebraic operation performed, rather than the problem in which it occurs.

Overall, these results suggested that transition-level embeddings capture transition semantics that generalize across problems, whereas state-level embeddings are dominated by problem-specific structure and formulation. This analysis motivates the use of transition-based sequences when the goal is to model student problem solving strategy rather than the structure and identity of problems. Hence, for the remaining analysis, we focused on learning representations for the entire solution pathway using transition-based sequences. 



\subsection{RQ2: Do sequence-level embeddings encode solution strategies?}
Table~\ref{tab:probe_results} reports the performance of probes on sequence-level embeddings across three evaluation tasks: Action Content prediction, Efficiency classification, and Action Sequence reconstruction. 
Among the frozen pretrained baselines, the mean-pooled embeddings from both MathBERT and MathBERT-mamut achieved similar results and were consistently better than math\_pretrained\_BERT. Overall, MathBERT-mamut was marginally better, reaching 96.0\% Micro-F1 and 91.7\% Macro-F1 on efficiency classification and perplexity of 2.34 on action sequence reconstruction. 

Models using a frozen encoder combined with our SimCSE sequence embeddings showed consistent improvements in action content prediction and action sequence reconstruction relative to the baselines. For instance, our encoder with sequence embeddings derived from MathBERT-mamut achieved the highest scores on the action content task (86.1\% Micro-F1 and 21.8\% Macro-F1), and action sequence reconstruction (2.06 perplexity and 77.0\% token-level accuracy). Performance on the efficiency classification was below the baseline models, which suggests that the learned sequence embeddings primarily capture information about transitions and implied actions rather than the efficiency of the solution sequence. 

Additionally, across different configurations of our SimCSE-based encoder—including the choice of frozen BERT model, the number of transformer layers ($L$), and the embedding dimension ($k$)—models with $L=1$ and $k=768$ consistently achieved the best performance. This suggests that smaller embedding dimensions substantially reduce representational capacity. In contrast, increasing the number of transformer layers may wash out token-level distinctions, optimizing sequence embeddings primarily for retrieval and yielding less discriminative representations.

Overall, incorporating the SimCSE sequence encoder improves both classification and reconstruction tasks relative to our baselines, with the largest gains observed in action content prediction and action sequence reconstruction. This suggests that the learned sequence embeddings capture richer information regarding solution strategies than the mean-pooled embeddings from pretrained BERT models.

\begin{table*}[!h]
\centering
\caption{Probe performance on sequence embeddings across evaluation tasks. Action content prediction is formulated as a multilabel classification task, Efficiency prediction as a three-class classification task, and Action Sequence evaluation reports reconstruction metrics.}
\begin{tabular}{lccccccc}
\toprule
\multirow{2}{*}{Model} 
& \multicolumn{2}{c}{Action Content} 
& \multicolumn{2}{c}{Efficiency Classification} 
& \multicolumn{2}{c}{Action Sequence} \\
\cmidrule(lr){2-3} \cmidrule(lr){4-5} \cmidrule(lr){6-7}
 & Micro-F1 & Macro-F1 & Micro-F1 & Macro-F1 & Perplexity & Accuracy (Token) \\
 \midrule
 \multicolumn{7}{l}{\textbf{Baselines (Mean Pool of Embeddings)}} \\
\midrule
math\_pretrained\_BERT (avg)       & 80.1 & 17.5 & 95.3 & 89.4 & 2.41 & 70.9 \\
MathBERT (avg)                     & 81.4 & 18.4 & \textbf{96.0} & 90.7 & 2.38 & 71.2 \\
MathBERT-mamut (avg)              & 80.9 & 18.2 & \textbf{96.0} & \textbf{91.7} & 2.34 & 71.6 \\
\midrule
\multicolumn{7}{l}{\textbf{Frozen encoder + our SimCSE sequence encoder}} \\
\midrule
math\_pretrained\_BERT             & 83.5 & 20.8 & 90.4 & 83.2 & 2.09 & 76.4 \\
\multicolumn{7}{l}{\hspace{2em}\textit{($L=1$, $k=768$)}} \\
MathBERT                           & 84.7 & 21.4 & 91.0 & 84.1 & 2.12 & 76.2 \\
\multicolumn{7}{l}{\hspace{2em}\textit{($L=1$, $k=768$)}} \\
MathBERT-mamut                    & \textbf{86.1} & \textbf{21.8} & 88.9 & 82.2 & \textbf{2.06} & \textbf{77.0} \\
\multicolumn{7}{l}{\hspace{2em}\textit{($L=1$, $k=768$)}} \\
\bottomrule
\end{tabular}
\label{tab:probe_results}
\end{table*}

\subsection{RQ3: How do our Measures of Mathematical Creativity Relate to Learning Outcomes?}\label{section-rq3}
We used the best performing sequence encoder configuration (i.e., MathBERT-Mamut, $k=768$, $L=1$) to compute the uniqueness, diversity, and conformity of students’ solution pathways in the FH2T game. Across all three short-term learning outcomes (conceptual, procedural, and flexibility scores), the uniqueness, diversity, and conformity of the employed strategies were all positive and statistically significant predictors (all $p < .001$). Among these, conformity, i.e., having strategies that are similar to the common and optimal pathway, showed the largest effect across outcomes. Prior knowledge, as measured by the pre-test score in the corresponding facet of algebraic learning, was also a strong predictor of short-term outcomes (pre-conceptual $b = 0.66$, $p < .001$; pre-procedural $b = 0.294$, $p < .001$; pre-flexibility $b = 0.277$, $p < .001$).

For distal learning outcomes, all three creativity and divergent thinking proxies again showed positive, statistically significant associations with 7th-grade state test scores. Similar to short-term learning, conformity exhibited the largest effect ($b = 10.3$, $p < .001$), followed by uniqueness ($b = 5.55$, $p < .001$) and diversity ($b = 4.97$, $p < .001$). The 5th Grade Test score was the strongest predictor ($b=50.43$, $p < .001$).

\begin{table*}[!ht]
\centering
\caption{Regression results showing associations between standardized predictors and learning outcomes. Predictor variables are standardized, while outcome variables are retained in their original scale to aid interpretability.}
\begin{tabular}{lcccccc|cc}
\toprule
\multirow{2}{*}{Predictor} 
& \multicolumn{6}{c|}{Short-term learning} 
& \multicolumn{2}{c}{Distal learning} \\
\cline{2-9}
& \multicolumn{2}{c}{Post-Conceptual Score\rule{0pt}{2.5ex}} 
& \multicolumn{2}{c}{Post-Procedural Score \rule{0pt}{2.5ex}} 
& \multicolumn{2}{c|}{Post-Flexibility Score \rule{0pt}{2.5ex}} 
& \multicolumn{2}{c}{Grade 7 State Score \rule{0pt}{2.5ex}} \\
\cline{2-9}
& $\boldsymbol{b}$ \rule{0pt}{2.5ex} & \textbf{SE} \rule{0pt}{2.5ex} & $\boldsymbol{b}$  \rule{0pt}{2.5ex}& \textbf{SE} \rule{0pt}{2.5ex}& $\boldsymbol{b}$ \rule{0pt}{2.5ex}& \textbf{SE} \rule{0pt}{2.5ex}& $\boldsymbol{b}$ \rule{0pt}{2.5ex}& \textbf{SE} \rule{0pt}{2.5ex}\\
\hline
Intercept & 1.890***& 0.038 & 1.422*** & 0.032 & 1.190*** & 0.031 & 560.1*** & 1.245 \\
Strat. Uniqueness & 0.185*** & 0.042 & 0.252*** & 0.036 & 0.175*** & 0.034 & 5.548*** & 1.368 \\
Strat. Diversity & 0.170*** & 0.038 & 0.116*** & 0.033 & 0.203*** & 0.032 & 4.976*** & 1.364 \\
Strat. Conformity  & 0.371*** & 0.047 & 0.363*** & 0.038 & 0.356*** & 0.037 & 10.32*** & 1.632 \\
Pre-Conceptual Score  & 0.661*** & 0.043 & -- & -- & -- & -- & -- & -- \\
Pre-Procedural Score & -- & -- & 0.294*** & 0.034 & -- & -- & -- & -- \\
Pre-Flexibility Score & -- & -- & -- & -- & 0.277*** & 0.033 & -- & -- \\
Grade 5 State Score & -- & -- & -- & -- & -- & -- & 50.426*** & 1.427 \\
\midrule
\textbf{Model fit} 
& \multicolumn{2}{c}{\shortstack{$N$ = 778 \\ $R^2$ = 0.42}}
& \multicolumn{2}{c}{\shortstack{$N$ = 778 \\ $R^2$ = 0.27}}
& \multicolumn{2}{c}{\shortstack{$N$ = 778 \\ $R^2$ = 0.27}}
& \multicolumn{2}{c}{\shortstack{$N$ = 766 \\ $R^2$ = 0.72}} \\
\bottomrule
\end{tabular}
\label{tab:regression_results}
\vspace{2pt}
\begin{minipage}{\textwidth}
\footnotesize
\textit{Note.} * $p<.05$, ** $p<.01$, *** $p<.001$.
\end{minipage}
\end{table*}

\section{Discussion}

A core objective of Educational Data Mining and Learning Analytics is to develop scalable and generalizable representations that capture student learning behaviors and strategies across tasks and platforms. This study contributes to this goal by examining whether sequence-level representations of student solution pathways can capture meaningful problem-solving strategy information beyond problem-specific structural features and formulations.

\paragraph{RQ1: Representing Mathematical Transitions}
In RQ1, we analyzed the geometric properties of individual state and transition embeddings. The results indicated that transition-based representations provided an effective abstraction for modeling mathematical problem-solving behavior. Importantly, these representations can be learned directly from pretrained language models without explicitly imposing a translation-based objective (e.g., TransE). This allows the model to leverage rich pretrained semantic structure while focusing on sequences of mathematical transformations rather than surface-level problem characteristics. Furthermore, this representation enables comparisons of student strategies across different problems and, potentially, across platforms, without relying on system-specific action labels or manual human coding.

\paragraph{RQ2: Probing Sequence-Level Embeddings for Strategy Information}
In RQ2, we evaluated the representational capacity of sequence embeddings derived from our trained SimCSE-based encoders. It is worth reiterating that the action labels were not used for training, instead we directly use transition-based representations derived from the mathematical states in the sequences. The action labels were used solely for evaluation purposes. 

We investigated the strategy information contained in the sequence-level embeddings using three probing tasks: action content detection, efficiency classification, and action-sequence reconstruction. These probes also served as a mechanism for selecting model hyperparameters for subsequent analysis. Results indicated that across encoder configurations, our SimCSE-based encoder model with a single transformer layer ($L=1$), embedding dimension $k=768$, using state embeddings from the frozen MathBERT-matmut encoder, consistently yielded the strongest overall performance.

Our model trained with this configuration outperformed baseline averaging methods on action content prediction and action-sequence reconstruction. This suggests that contrastive training encourages sequence-level embeddings to encode the mathematical transitions and action orderings. However, performance on efficiency prediction did not consistently improve under SimCSE training. One plausible explanation is that the contrastive objective, particularly when paired with in-batch hard negatives, may push apart sequences with similar efficiency while prioritizing similarity in action composition and order. This reflects an inherent tradeoff in our representation learning objectives. 


An additional finding is that increasing the number of transformer layers in the sequence encoder degraded downstream probe performance. One plausible explanation is our choice of model architecture: the model employs mean pooling over the final transformer layer, while prior work suggests that universal and general sequence embeddings benefit from aggregating information across layers ~\cite{ma_universal_2019, gao_simcse_2021}, such as the first and last, rather than relying on the deepest representations alone. As depth increases, the final layer may increasingly encode features optimized for the retrieval objective, rather than general semantic structure and solution strategy. Consequently, greater model capacity does not translate to better probe performance and may instead induce less transferable representations.

\paragraph{RQ3: Relation of Creativity and Divergent Thinking Measures to Learning Outcomes}
For RQ3, we calculated the uniqueness, diversity, and conformity of students' problem-solving pathways using the best-performing model from our probe analysis. The results from the regression analysis in RQ3 (Section~\ref{section-rq3}) indicated that the proposed strategy-based measures of creativity and divergent thinking are strongly associated with both short-term and distal learning outcomes (Table~\ref{tab:regression_results}). Across all short-term outcomes--conceptual, procedural, and flexibility scores--strategy uniqueness, diversity, and conformity were each positively and statistically significant predictors, even after accounting for prior knowledge. Similarly, for distal learning outcomes, all three measures showed positive and statistically significant associations with 7th-grade state test scores after accounting for 5th-grade state test performance. 

Among the three measures, strategy conformity exhibited the strongest association across all learning outcomes. In the context of this study, conformity to a common optimal strategy may reflect a goal-oriented problem-solving approach where students aim to arrive at a single correct and optimal solution by evaluating as few ideas or solutions as possible \cite{dehaan_teaching_2009}. However, this does not imply an absence of exploration or divergent thinking in students, but rather that they may have ultimately converged on a commonly used optimal strategy during problem solving.

At the same time, the positive associations observed for strategy uniqueness and diversity suggested that additional benefits may arise from exploring a wider range and more unique strategies. This ability to move beyond fixed and conventional strategies to generate multiple solution approaches may be indicative of mathematical creativity \cite{haylock_mathematical_1987}. Additionally, taken together, the positive associations observed across the three measures suggest that effective problem solving may involve both divergent and convergent thinking. Prior work in mathematical creativity supports this interpretation, where creative problem solving is characterized as generating a diverse set of strategies and subsequently refining them to arrive at an optimal solution \cite{sternberg_concept_1999}.

Overall, these findings provided preliminary evidence that similarity- and distance-based measures derived from sequence-level embeddings representing students' problem-solving strategy are meaningfully related to learning outcomes across both short-term and distal time frames, complementing traditional performance and achievement measures. 

\section{Limitations}
While our approach enables strategy-level analysis that generalizes across problems, its applicability is constrained by several factors. First, we considered a limited set of base encoders, which restricts the generality of our findings across substantially different model architectures. Additionally, the inclusion of domain-specific encoders (e.g., Forte encoder~\cite{wang_scientific_2021}), that focus on modeling expressions rather than the expression and surrounding context, may lead to better sequence representations. Secondly, we explored only simple difference-based transition embeddings to represent the solution sequence. Using more expressive transition functions (e.g., $(h_t, h_{t-1}, h_t \odot h_{t-1}, |h_t - h_{t-1}|)$) may capture richer relational structures. Third, our analysis relied on a subset of data from the RCT and focused on the FH2T platform. Larger and more diverse datasets are necessary to train sequence encoders that better capture student problem-solving strategies. Finally, our evaluation relied primarily on probing tasks to assess representational capacity. While the probes provided useful insights into the information contained in the embeddings, establishing construct validity for creativity-related measures ultimately requires comparison against human-labeled datasets.

\section{Future Directions}

This work opens several directions for future research for enhancing teaching and learning. One promising application is the development of systems that automatically suggest alternative solution approaches that are maximally different from a student’s current strategy, encouraging exploration and divergent thinking. More refined and theory-grounded measures of mathematical creativity could also be developed using the proposed representations. Such measures could support instructional interventions by encouraging students to try multiple strategies, guiding the design of problems that require strategic diversity, and providing teachers with interpretable metrics of students’ strategic flexibility. These insights could ultimately help educators adapt instruction to foster mathematical flexibility, creativity, and divergent thinking.

\section{Conclusion}
In this study, we introduced a method for learning representations of entire algebraic solution pathways. By leveraging pretrained mathematical encoders and contrastive sequence-level learning via SimCSE, we showed that transition-based embeddings capture the structural and strategic characteristics of student solutions. Evaluation across action-type content, solution efficiency, and sequence reconstruction tasks indicates that these embeddings encode meaningful information about strategies and sequence order. Moreover, embedding-based measures of strategy uniqueness, diversity, and conformity were strongly associated with both short-term and long-term learning outcomes, suggesting that pathway-level representations can serve as scalable proxies for mathematical creativity and divergent thinking. This work advances the analysis of student problem-solving by providing a scalable, platform-agnostic method for quantifying the diversity and uniqueness of algebraic solution strategies.

\section{Acknowledgments}
The research reported here was supported by the Institute of Education Sciences, U.S. Department of Education, through an Efficacy and Replication Grant (R305A180401) and an NSF CAREER Grant (2142984) to Worcester Polytechnic Institute. The opinions expressed are those of the authors and do not represent the views of the Institute or the U.S. Department of Education.

%
\bibliographystyle{abbrv}
\bibliography{sigproc}  

\begin{thebibliography}{10}

\bibitem{adi_fine-grained_2017}
Y.~Adi, E.~Kermany, Y.~Belinkov, O.~Lavi, and Y.~Goldberg.
\newblock Fine-grained {Analysis} of {Sentence} {Embeddings} {Using} {Auxiliary} {Prediction} {Tasks}, Feb. 2017.
\newblock arXiv:1608.04207 [cs].

\bibitem{agirre2016semeval}
E.~Agirre, C.~Banea, D.~Cer, M.~Diab, A.~Gonzalez-Agirre, R.~Mihalcea, G.~Rigau, and J.~Wiebe.
\newblock Semeval-2016 task 1: {Semantic} textual similarity, monolingual and cross-lingual evaluation.
\newblock In {\em Proceedings of the 10th international workshop on semantic evaluation ({SemEval}-2016)}, pages 497--511, 2016.

\bibitem{alfieri2013learning}
L.~Alfieri, T.~J. Nokes-Malach, and C.~D. Schunn.
\newblock Learning through case comparisons: {A} meta-analytic review.
\newblock {\em Educational Psychologist}, 48(2):87--113, 2013.

\bibitem{baker2006human}
R.~S. Baker, A.~T. Corbett, and A.~Z. Wagner.
\newblock Human classification of low-fidelity replays of student actions.
\newblock In {\em Proceedings of the educational data mining workshop at the 8th international conference on intelligent tutoring systems}, volume 2002, pages 29--36, 2006.

\bibitem{beaty_automating_2021}
R.~E. Beaty and D.~R. Johnson.
\newblock Automating creativity assessment with {SemDis}: {An} open platform for computing semantic distance.
\newblock {\em Behavior Research Methods}, 53(2):757--780, 2021.

\bibitem{bordes_translating_2013}
A.~Bordes, N.~Usunier, A.~Garcia-Duran, J.~Weston, and O.~Yakhnenko.
\newblock Translating {Embeddings} for {Modeling} {Multi}-relational {Data}.
\newblock In {\em Advances in {Neural} {Information} {Processing} {Systems}}, volume~26. Curran Associates, Inc., 2013.

\bibitem{chan_slow_2022}
J.~Y.-C. Chan, E.~R. Ottmar, and J.-E. Lee.
\newblock Slow down to speed up: {Longer} pause time before solving problems relates to higher strategy efficiency.
\newblock {\em Learning and Individual Differences}, 93:102109, 2022.

\bibitem{conneau_senteval_2018}
A.~Conneau and D.~Kiela.
\newblock {SentEval}: {An} {Evaluation} {Toolkit} for {Universal} {Sentence} {Representations}, Mar. 2018.

\bibitem{decker-woodrow_impacts_2023}
L.~E. Decker-Woodrow, C.~A. Mason, J.-E. Lee, J.~Y.-C. Chan, A.~Sales, A.~Liu, and S.~Tu.
\newblock The impacts of three educational technologies on algebraic understanding in the context of {COVID}-19.
\newblock {\em AERA open}, 9:23328584231165919, 2023.
\newblock Publisher: SAGE Publications Sage CA: Los Angeles, CA.

\bibitem{dehaan_teaching_2009}
R.~L. DeHaan.
\newblock Teaching creativity and inventive problem solving in science.
\newblock {\em CBE life sciences education}, 8(3):172--181, 2009.

\bibitem{drechsel_mamut_2025}
J.~Drechsel, A.~Reusch, and S.~Herbold.
\newblock {MAMUT}: {A} {Novel} {Framework} for {Modifying} {Mathematical} {Formulas} for the {Generation} of {Specialized} {Datasets} for {Language} {Model} {Training}, July 2025.
\newblock arXiv:2502.20855 [cs].

\bibitem{feldman_automatic_2018}
M.~Q. Feldman, J.~Y. Cho, M.~Ong, S.~Gulwani, Z.~Popović, and E.~Andersen.
\newblock Automatic {Diagnosis} of {Students}' {Misconceptions} in {K}-8 {Mathematics}.
\newblock In {\em Proceedings of the 2018 {CHI} {Conference} on {Human} {Factors} in {Computing} {Systems}}, {CHI} '18, pages 1--12, New York, NY, USA, Apr. 2018. Association for Computing Machinery.

\bibitem{gao_simcse_2021}
T.~Gao, X.~Yao, and D.~Chen.
\newblock {SimCSE}: {Simple} {Contrastive} {Learning} of {Sentence} {Embeddings}.
\newblock In M.-F. Moens, X.~Huang, L.~Specia, and S.~W.-t. Yih, editors, {\em Proceedings of the 2021 {Conference} on {Empirical} {Methods} in {Natural} {Language} {Processing}}, pages 6894--6910, Online and Punta Cana, Dominican Republic, Nov. 2021. Association for Computational Linguistics.

\bibitem{grover_computational_2013}
S.~Grover and R.~Pea.
\newblock Computational {Thinking} in {K}–12: {A} {Review} of the {State} of the {Field}.
\newblock {\em Educational Researcher}, 42(1):38--43, Jan. 2013.

\bibitem{haylock_mathematical_1987}
D.~W. Haylock.
\newblock Mathematical creativity in schoolchildren.
\newblock {\em The Journal of Creative Behavior}, 21(1):48--59, 1987.

\bibitem{hendrycks_gaussian_2023}
D.~Hendrycks and K.~Gimpel.
\newblock Gaussian {Error} {Linear} {Units} ({GELUs}), June 2023.
\newblock arXiv:1606.08415 [cs].

\bibitem{hill_learning_2016}
F.~Hill, K.~Cho, and A.~Korhonen.
\newblock Learning {Distributed} {Representations} of {Sentences} from {Unlabelled} {Data}.
\newblock In K.~Knight, A.~Nenkova, and O.~Rambow, editors, {\em Proceedings of the 2016 {Conference} of the {North} {American} {Chapter} of the {Association} for {Computational} {Linguistics}: {Human} {Language} {Technologies}}, pages 1367--1377, San Diego, California, June 2016. Association for Computational Linguistics.

\bibitem{hochreiter_long_1997}
S.~Hochreiter and J.~Schmidhuber.
\newblock Long {Short}-{Term} {Memory}.
\newblock {\em Neural Computation}, 9(8):1735--1780, Nov. 1997.

\bibitem{hong_systematic_2023}
W.~Hong, J.~R. Star, R.-D. Liu, R.~Jiang, and X.~Fu.
\newblock A {Systematic} {Review} of {Mathematical} {Flexibility}: {Concepts}, {Measurements}, and {Related} {Research}.
\newblock {\em Educational Psychology Review}, 35(4):104, Dec. 2023.

\bibitem{kennard2016evaluating}
N.~N. Kennard, G.~Angeli, and C.~D. Manning.
\newblock Evaluating word embeddings using a representative suite of practical tasks.
\newblock In {\em Proceedings of the 1st workshop on evaluating vector-space representations for nlp}, pages 19--23, 2016.

\bibitem{knuth_does_2006}
E.~J. Knuth, A.~C. Stephens, N.~M. McNeil, and M.~W. Alibali.
\newblock Does understanding the equal sign matter? {Evidence} from solving equations.
\newblock {\em Journal for research in Mathematics Education}, 37(4):297--312, 2006.

\bibitem{koedinger_real_2004}
K.~R. Koedinger and M.~J. Nathan.
\newblock The {Real} {Story} {Behind} {Story} {Problems}: {Effects} of {Representations} on {Quantitative} {Reasoning}.
\newblock {\em Journal of the Learning Sciences}, 13(2):129--164, Apr. 2004.
\newblock \_eprint: https://doi.org/10.1207/s15327809jls1302\_1.

\bibitem{kolb_generalizing_2019}
J.~Kolb, S.~Farrar, and Z.~A. Pardos.
\newblock Generalizing {Expert} {Misconception} {Diagnoses} through {Common} {Wrong} {Answer} {Embedding}.
\newblock Technical report, International Educational Data Mining Society, July 2019.
\newblock ERIC Number: ED599212.

\bibitem{kwon_cultivating_2006}
O.~N. Kwon, J.~H. Park, and J.~S. Park.
\newblock Cultivating divergent thinking in mathematics through an open-ended approach.
\newblock {\em Asia Pacific Education Review}, 7(1):51--61, July 2006.

\bibitem{lee_does_2022}
J.-E. Lee, J.~Y.-C. Chan, A.~Botelho, and E.~Ottmar.
\newblock Does slow and steady win the race?: {Clustering} patterns of students’ behaviors in an interactive online mathematics game.
\newblock {\em Educational technology research and development}, 70(5):1575--1599, 2022.

\bibitem{lee_perceptual_2022}
J.-E. Lee, C.~B. Hornburg, J.~Y.-C. Chan, and E.~Ottmar.
\newblock Perceptual and {Number} {Effects} on {Students}’ {Initial} {Solution} {Strategies} in an {Interactive} {Online} {Mathematics} {Game}.
\newblock {\em Journal of Numerical Cognition}, 8(1):166--182, Mar. 2022.

\bibitem{leikin_exploring_2009}
R.~Leikin.
\newblock Exploring {Mathematical} {Creativity} {Using} {Multiple} {Solution} {Tasks}.
\newblock In {\em Creativity in {Mathematics} and the {Education} of {Gifted} {Students}}, pages 129--145. Brill, Jan. 2009.
\newblock Section: Creativity in Mathematics and the Education of Gifted Students.

\bibitem{leikin2007multiple}
R.~Leikin and M.~Lev.
\newblock Multiple solution tasks as a magnifying glass for observation of mathematical creativity.
\newblock In {\em Proceedings of the 31st international conference for the psychology of mathematics education}, volume~3, pages 161--168, 2007.

\bibitem{li_sentence_2023}
H.~Li, M.~Xu, and Y.~Song.
\newblock Sentence {Embedding} {Leaks} {More} {Information} than {You} {Expect}: {Generative} {Embedding} {Inversion} {Attack} to {Recover} the {Whole} {Sentence}.
\newblock In A.~Rogers, J.~Boyd-Graber, and N.~Okazaki, editors, {\em Findings of the {Association} for {Computational} {Linguistics}: {ACL} 2023}, pages 14022--14040, Toronto, Canada, July 2023. Association for Computational Linguistics.

\bibitem{logeswaran_efficient_2018}
L.~Logeswaran and H.~Lee.
\newblock An efficient framework for learning sentence representations.
\newblock Feb. 2018.

\bibitem{ma_universal_2019}
X.~Ma, Z.~Wang, P.~Ng, R.~Nallapati, and B.~Xiang.
\newblock Universal {Text} {Representation} from {BERT}: {An} {Empirical} {Study}, Oct. 2019.
\newblock arXiv:1910.07973 [cs].

\bibitem{maaten2008visualizing}
L.~v.~d. Maaten and G.~Hinton.
\newblock Visualizing data using t-{SNE}.
\newblock {\em Journal of machine learning research}, 9(Nov):2579--2605, 2008.

\bibitem{morris_text_2023}
J.~X. Morris, V.~Kuleshov, V.~Shmatikov, and A.~M. Rush.
\newblock Text {Embeddings} {Reveal} ({Almost}) {As} {Much} {As} {Text}, Oct. 2023.
\newblock arXiv:2310.06816 [cs].

\bibitem{ottmar_teaching_2012}
E.~Ottmar, D.~Landy, and R.~Goldstone.
\newblock Teaching the perceptual structure of algebraic expressions: {Preliminary} findings from the pushing symbols intervention.
\newblock In {\em Proceedings of the {Annual} {Meeting} of the {Cognitive} {Science} {Society}}, volume~34, 2012.

\bibitem{ottmar_data_2023}
E.~Ottmar, J.-E. Lee, K.~Vanacore, S.~Pradhan, L.~Decker-Woodrow, and C.~A. Mason.
\newblock Data from the efficacy study of from here to there! {A} dynamic technology for improving algebraic understanding.
\newblock {\em Journal of Open Psychology Data}, 11(1), 2023.

\bibitem{ottmar_getting_2015}
E.~R. Ottmar, D.~Landy, R.~Goldstone, and E.~Weitnauer.
\newblock Getting {From} {Here} to {There}! : {Testing} the {Effectiveness} of an {Interactive} {Mathematics} {Intervention} {Embedding} {Perceptual} {Learning}.
\newblock {\em Proceedings of the Annual Meeting of the Cognitive Science Society}, 37(0), 2015.

\bibitem{paquette2014reengineering}
L.~Paquette, A.~de~Carvahlo, R.~Baker, and J.~Ocumpaugh.
\newblock Reengineering the feature distillation process: {A} case study in detection of gaming the system.
\newblock In {\em Educational data mining 2014}, 2014.

\bibitem{pradhan_gamification_2024}
S.~Pradhan, A.~Gurung, and E.~Ottmar.
\newblock Gamification and deadending: {Unpacking} performance impacts in algebraic learning.
\newblock In {\em Proceedings of the 14th learning analytics and knowledge conference}, pages 899--906, 2024.

\bibitem{pradhan2025mathflowlens}
S.~Pradhan, E.~Ottmar, A.~Gurung, and J.-E. Lee.
\newblock {MathFlowLens}: a classification and visualization tool for analyzing students’ procedural pathways: {S}. {Pradhan} et al.
\newblock {\em Educational technology research and development}, pages 1--26, 2025.

\bibitem{reimers_sentence-bert_2019}
N.~Reimers and I.~Gurevych.
\newblock Sentence-{BERT}: {Sentence} {Embeddings} using {Siamese} {BERT}-{Networks}.
\newblock In K.~Inui, J.~Jiang, V.~Ng, and X.~Wan, editors, {\em Proceedings of the 2019 {Conference} on {Empirical} {Methods} in {Natural} {Language} {Processing} and the 9th {International} {Joint} {Conference} on {Natural} {Language} {Processing} ({EMNLP}-{IJCNLP})}, pages 3982--3992, Hong Kong, China, Nov. 2019. Association for Computational Linguistics.

\bibitem{reusch_investigating_2024}
A.~Reusch, J.~Gonsior, C.~Hartmann, and W.~Lehner.
\newblock Investigating the {Usage} of {Formulae} in {Mathematical} {Answer} {Retrieval}.
\newblock In {\em Advances in {Information} {Retrieval}: 46th {European} {Conference} on {Information} {Retrieval}, {ECIR} 2024, {Glasgow}, {UK}, {March} 24–28, 2024, {Proceedings}, {Part} {I}}, pages 247--261, Berlin, Heidelberg, Mar. 2024. Springer-Verlag.

\bibitem{reusch2022transformer}
A.~Reusch, M.~Thiele, and W.~Lehner.
\newblock Transformer-encoder and decoder models for questions on math.
\newblock 2022.
\newblock Authority: CLEF.

\bibitem{ritter_identifying_2019}
S.~Ritter, R.~Baker, V.~Rus, and G.~Biswas.
\newblock Identifying strategies in student problem solving.
\newblock {\em Design Recommendations for Intelligent Tutoring Systems}, 7:59--70, 2019.

\bibitem{rittle-johnson_not_2015}
B.~Rittle-Johnson, M.~Schneider, and J.~R. Star.
\newblock Not a {One}-{Way} {Street}: {Bidirectional} {Relations} {Between} {Procedural} and {Conceptual} {Knowledge} of {Mathematics}.
\newblock {\em Educational Psychology Review}, 27(4):587--597, Dec. 2015.

\bibitem{rittle-johnson_developing_2001}
B.~Rittle-Johnson, R.~S. Siegler, and M.~W. Alibali.
\newblock Developing conceptual understanding and procedural skill in mathematics: {An} iterative process.
\newblock {\em Journal of educational psychology}, 93(2):346, 2001.

\bibitem{rittlejohnson_developing_2012}
B.~Rittle‐Johnson, J.~R. Star, and K.~Durkin.
\newblock Developing procedural flexibility: {Are} novices prepared to learn from comparing procedures?
\newblock {\em British Journal of Educational Psychology}, 82(3):436--455, Sept. 2012.

\bibitem{schneider_relations_2011}
M.~Schneider, B.~Rittle-Johnson, and J.~R. Star.
\newblock Relations among conceptual knowledge, procedural knowledge, and procedural flexibility in two samples differing in prior knowledge.
\newblock {\em Developmental psychology}, 47(6):1525, 2011.

\bibitem{schoenfeld_problem_2007}
A.~H. Schoenfeld.
\newblock Problem solving in the {United} {States}, 1970–2008: research and theory, practice and politics.
\newblock {\em ZDM}, 39(5-6):537--551, Sept. 2007.

\bibitem{selden2009reflections}
A.~Selden and J.~Selden.
\newblock Reflections on foundations for success: {The} final report of the national mathematics advisory panel, 2009.

\bibitem{shakya_student_2021}
A.~Shakya, V.~Rus, and D.~Venugopal.
\newblock Student {Strategy} {Prediction} {Using} a {Neuro}-{Symbolic} {Approach}.
\newblock Technical report, International Educational Data Mining Society, 2021.
\newblock ERIC Number: ED615630.

\bibitem{shen_mathbert_2023}
J.~T. Shen, M.~Yamashita, E.~Prihar, N.~Heffernan, X.~Wu, B.~Graff, and D.~Lee.
\newblock {MathBERT}: {A} {Pre}-trained {Language} {Model} for {General} {NLP} {Tasks} in {Mathematics} {Education}, Aug. 2023.
\newblock arXiv:2106.07340 [cs].

\bibitem{srivastava_dropout_2014}
N.~Srivastava, G.~Hinton, A.~Krizhevsky, I.~Sutskever, and R.~Salakhutdinov.
\newblock Dropout: a simple way to prevent neural networks from overfitting.
\newblock {\em J. Mach. Learn. Res.}, 15(1):1929--1958, Jan. 2014.

\bibitem{stanton2021fostering}
J.~D. Stanton, A.~J. Sebesta, and J.~Dunlosky.
\newblock Fostering metacognition to support student learning and performance.
\newblock {\em CBE—Life Sciences Education}, 20(2):fe3, 2021.

\bibitem{sternberg_concept_1999}
R.~J. Sternberg and T.~I. Lubart.
\newblock The concept of creativity: {Prospects} and paradigms.
\newblock In {\em Handbook of creativity}, pages 3--15. Cambridge University Press, New York, NY, US, 1999.

\bibitem{tabach_algebraic_2017}
M.~Tabach and A.~Friedlander.
\newblock Algebraic procedures and creative thinking.
\newblock {\em ZDM}, 49(1):53--63, Mar. 2017.

\bibitem{thapa_magar_learning_2024}
A.~Thapa~Magar, S.~E. Fancsali, V.~Rus, A.~Murphy, S.~Ritter, and D.~Venugopal.
\newblock Learning {Representations} for {Math} {Strategies} using {BERT}.
\newblock In {\em Proceedings of the {Eleventh} {ACM} {Conference} on {Learning} @ {Scale}}, L@{S} '24, pages 514--518, New York, NY, USA, July 2024. Association for Computing Machinery.

\bibitem{thapa_magar_can_2025}
A.~Thapa~Magar, A.~Shakya, S.~E. Fancsali, V.~Rus, A.~Murphy, S.~Ritter, and D.~Venugopal.
\newblock "{Can} {A} {Language} {Model} {Represent} {Math} {Strategies}?": {Learning} {Math} {Strategies} from {Big} {Data} using {BERT}.
\newblock In {\em Proceedings of the 15th {International} {Learning} {Analytics} and {Knowledge} {Conference}}, {LAK} '25, pages 655--666, New York, NY, USA, Mar. 2025. Association for Computing Machinery.

\bibitem{vaswani2017attention}
A.~Vaswani, N.~Shazeer, N.~Parmar, J.~Uszkoreit, L.~Jones, A.~N. Gomez, {\L}.~Kaiser, and I.~Polosukhin.
\newblock Attention is all you need.
\newblock {\em Advances in neural information processing systems}, 30, 2017.

\bibitem{verschaffel_conceptualizing_2009}
L.~Verschaffel, K.~Luwel, J.~Torbeyns, and W.~Van~Dooren.
\newblock Conceptualizing, investigating, and enhancing adaptive expertise in elementary mathematics education.
\newblock {\em European Journal of Psychology of Education}, 24(3):335--359, Sept. 2009.

\bibitem{wang_scientific_2021}
Z.~Wang, M.~Zhang, R.~G. Baraniuk, and A.~S. Lan.
\newblock Scientific {Formula} {Retrieval} via {Tree} {Embeddings}.
\newblock In {\em 2021 {IEEE} {International} {Conference} on {Big} {Data} ({Big} {Data})}, pages 1493--1503, Dec. 2021.

\bibitem{welder_improving_2012}
R.~M. Welder.
\newblock Improving {Algebra} {Preparation}: {Implications} {From} {Research} on {Student} {Misconceptions} and {Difficulties}.
\newblock {\em School Science and Mathematics}, 112(4):255--264, 2012.
\newblock \_eprint: https://onlinelibrary.wiley.com/doi/pdf/10.1111/j.1949-8594.2012.00136.x.

\bibitem{zhang_math_2021}
M.~Zhang, Z.~Wang, R.~Baraniuk, and A.~Lan.
\newblock Math {Operation} {Embeddings} for {Open}-ended {Solution} {Analysis} and {Feedback}, Apr. 2021.

\end{thebibliography}
%

\balancecolumns
\end{document}